\documentclass[11pt,twoside]{article}
\usepackage[dvips]{graphics}
\usepackage{newpasp}
\newcommand{\oversim}[2]{\protect{\mbox{\lower0.5ex\vbox{%
   \baselineskip=0pt\lineskip=0.2ex
   \ialign{$\mathsurround=0pt #1\hfil##\hfil$\crcr#2\crcr\sim\crcr}}}}} 
\newcommand{\simgreat}{\mbox{$\,\mathrel{\mathpalette\oversim>}\,$}} % >~ sign
\newcommand{\simless} {\mbox{$\,\mathrel{\mathpalette\oversim<}\,$}} % <~ sign
\def\edcomment#1{\iffalse\marginpar{\raggedright\sl#1\/}\else\relax\fi}
\reversemarginpar

\begin{document}
\title{Binary Stars in Young Clusters -- a Theoretical Perspective}
 \author{Pavel Kroupa} 

%\affil{Institut f\"ur Theoretische Physik und Astrophysik der Universit\"at
%Kiel\\ Olshausenstra{\ss}e 40, D-24118 Kiel, Germany}

\affil{Institut f\"ur Theoretische Astrophysik\\
Tiergartenstr.~15, D-69121 Heidelberg, Germany}

\begin{abstract}
The preponderance of binary systems in all known stellar populations
makes them exciting {\it dynamical agents} for research on topics as
varied as star formation, star-cluster dynamics and the interiors of
young and old stars. Today we know that the Galactic-field binary
population is probably a dynamically evolved version of the
Taurus--Auriga pre-main sequence population, and that the initial
distributions of binary-star orbital elements are probably universal.
Furthermore, $N$-body calculations tentatively suggest that OB stars
form in energetic binaries near cluster cores, and that binaries with
'forbidden' orbital elements that are produced in stellar encounters,
may turn out to be very useful windows into stellar interiors,
potentially allowing tests of pre-main sequence evolution theory as
well as of models of main-sequence stars.
\end{abstract}


%======================================================================
\section{Introduction}
\noindent
Most 'stars' in the sky are multiple systems, and observations
demonstrate that the binary proportion of 'isolated' pre-main sequence
stars is typically much higher than in the Galactic-field (GF) (Ghez
et al. 1997; K\"ohler \& Leinert 1998; Duchene 1999).  Differences
such as this contain clues about the origins of various
populations. The disruption of binary systems in their stellar birth
aggregates may naturally reduce the binary proportion to the GF level.
This is an exciting notion, opening the possibility of learning
something about the configuration of the birth aggregate typical for
present-day star formation (sf).

The study of the evolution of a primordial binary population in a
young cluster thus becomes important for understanding the origin of
the distribution of binary-star orbital elements. Cluster dynamics
with a substantial primordial binary population differs from that of
single-star clusters because the number of interacting systems changes
with time owing to binary disruption affecting the relaxation process,
and because binary systems contain additional degrees of freedom that
single stars do not. These internal degrees of freedom affect the
energy exchanges that occur during close encounters, changing the
energy spectrum of escaping stars and the energy budget of the whole
cluster, ultimately being able to arrest core collapse in massive star
clusters (e.g. Giannone \& Molteni 1985; Giersz \& Spurzem 2000).

This text addresses related questions, with the caveat that many
important and interesting problems remain to be studied.
Complementary discussions are available in Kroupa (2000a, KI; Kroupa
2000b, KII), this contribution (KIII) concentrating primarily on the
implications of stellar-dynamical interactions on the binary-star
orbital characteristics and high-velocity stars.

In the following, the mass, $m$, period, $P$ (always in days),
eccentricity, $e$, and mass-ratio, $q=m_2/m_1; m_2\le m_1$, of a
stellar system (a binary, or a single with no $P$, $e$, $q$) are
collectively referred to as its {\it dynamical properties}.  The
primordial, or birth, distribution functions of these quantities are
the initial mass function (IMF), and the initial period, eccentricity
and mass-ratio distributions (IPF, IEF, IMRF, respectively). The
distribution functions are written as, $f_x$, where $f_x\,dx$ is, for
example, the number of orbits with orbital parameter $x$ between $x$
and $x+dx$. The total (all periods and all primaries) binary
proportion is $f_{\rm tot} = N_{\rm bin}/\left(N_{\rm bin}+N_{\rm
sing}\right)$, with $N_{\rm bin}$ being the number of binaries and
$N_{\rm sing}$ the number of single stars in the sample.

There are two main mechanisms that evolve the dynamical properties:
{\it eigenevolution} through system-internal processes that
redistribute angular momentum and energy (e.g. tidal circularisation,
disk--companion-star interactions, outflows), and {\it stimulated
evolution} through encounters with neighbours. Eigenevolution affects
short-period systems ($P\simless 10^3$~days), whereas stimulated
evolution affects long-period systems.

One key goal of the work reported here is to find out if the IPF, IEF
and IMRF are universal, or if systematic variations with sf conditions
are evident in the available data (Durisen \& Sterzik 1994).

%======================================================================
\section{Origin of the Galactic Field Population}
\label{sec:idps}
\noindent
The difference between the period distribution of T~Tauri (TT)
binaries, $f_{\rm P}^{\rm TT}$, and of GF late-type binaries, $f_{\rm
P}^{\rm GF}$, ($f_{\rm P}^{\rm TT} \approx 2\,f_{\rm P}^{\rm GF}$ for
$P\simgreat 10^4$~days) as well as between the respective mass-ratio
distributions, $f_{\rm q}^{\rm TT}$ and $f_{\rm q}^{\rm GF}$, ($f_{\rm
q}^{\rm TT} > f_{\rm q}^{\rm GF}$ for $q\simless 0.35$, Leinert et
al. 1993) can readily be understood if most GF stars stem from
clusters that are {\it dynamically equivalent} to a cluster consisting
of $N_{\rm bin}=200$ binaries with a half-mass radius $R_{0.5}=0.8$~pc
(the {\it dominant mode cluster}).  This result is obtained by
performing $N$-body calculations of a library of clusters with
different properties (in this instance different $R_{0.5}$ but the
same $N_{\rm bin}=200$) (Kroupa 1995a, K1). The approach is called
{\it inverse dynamical population synthesis} (IDPS) because the
typical sf structure is inferred from the dynamical properties of GF
stars; {\it dynamical population synthesis} (DPS) being the
construction of a synthetic GF population from a distribution of
stellar birth aggregates (a future goal).

How IDPS works is demonstrated in figs.~4 and~5 of K1. The resulting
dominant-mode cluster, in which the pre-main sequence period and
mass-ratio distributions evolve to the observed GF distributions, is
remarkably similar to the most-common embedded cluster observed in the
Orion molecular cloud (Lada \& Lada 1991), which is also held to be
the dominant mode of sf.

This work shows that the discordant TT and GF populations can be
unified; the latter being a dynamically evolved version of the former.
With IDPS the typical sf structure can thus be identified if the
properties of primordial binary systems are assumed to be universally
similar to the Taurus--Auriga population.  Conversely, adopting
(retrospectively) the embedded clusters described by Lada \& Lada
(1991) as the typical sf events, it follows that the {\it IPF, IEF and
IMRF are remarkably universal}. This is supported by the period
distribution of binaries in the ONC and the Pleiades; both can be
viewed as being evolved versions of a Taurus--Auriga-like distribution
(Kroupa, Petr \& McCaughrean 1999; Kroupa, Aarseth \& Hurley 2000).

\begin{figure}
\plotfiddle{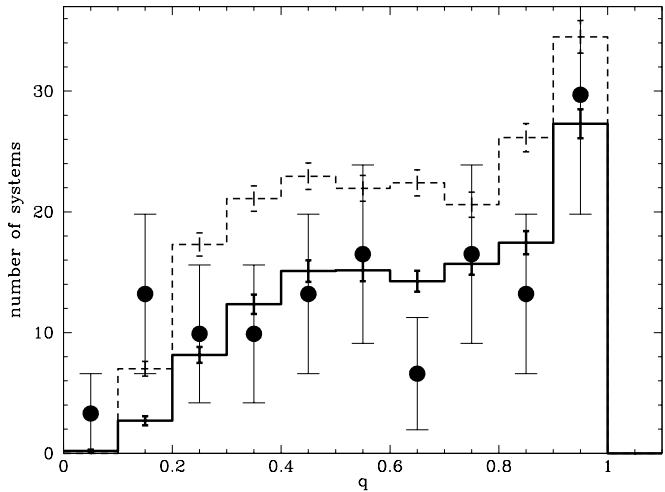}{6cm}{-90}{35}{35}{-150}{200}
\caption{\small The overall model GF mass-ratio distribution ($0.1\le
m_{1,2}\le 1.1\,M_\odot$) is the solid histogram, whereas the IMRF
(random pairing from the GF IMF) is shown as the dashed histogram
(from fig.12 in K2).  The peak at $q=1$ results from adjustment of
orbital parameters during pre-main sequence
eigenevolution. Observational data from Reid \& Gizis (1997) are shown
as solid dots, after removing WD companions and scaling to the
model. This 8~pc sample is not complete and may be biased towards
$q=1$ systems (Henry et al. 1997). Nevertheless, the agreement between
model and data is striking. }
\label{fig:qall}
\end{figure}

A more realistic primordial binary population by Kroupa (1995b, K2),
including the entire range of orbital periods and a model for
eigenevolution together with many more $N$-body computations, shows
that the GF population can be reproduced beautifully, if most stars
form in the above mentioned dominant-mode cluster
(e.g. Fig.~\ref{fig:qall}).
%-------------------------------------
% Figs: NB_EVAL/Figures/q_final_iau200.sm
%-------------------------------------
An elaborate study of this model GF population confirms that
essentially all triple and quadruple systems in the GF must be
primordial, that the dependence of the model binary proportion on
spectral type reproduces the observations, and shows that the observed
specific angular momentum distribution, $f_{\rm J}$, of molecular
cloud cores is a smooth extension to large values of the primordial
binary-system $f_{\rm J}$.

%======================================================================
\section{The Binary Star -- Cluster Connection: Fundamentals}
\label{sec:bincl}

{\bf Mass segregation and massive sub-system:} Massive systems sink
towards the centre through energy equipartition, thereby gaining
potential energy. This {\it heats} the cluster which expands.  Once
the massive stars assemble near the cluster core they interact,
exchanging companions for more massive contemporaries, ejecting the
divorced partners, and being ejected themselves when one of the
involved systems hardens sufficiently (for more details see KII).

\noindent
{\bf Hard and soft binaries:} Further energy exchanges involve {\it
soft} and {\it hard} binaries.  A soft binary has a binding energy,
$\left|e_{\rm b}\right| \ll e_{\rm k}$, where $e_{\rm k}$ is the
average kinetic energy of systems in the cluster. For a hard binary,
$\left|e_{\rm b}\right| \gg e_{\rm k}$.  The average energy transfer
can be understood with the equipartition argument: Let the orbital
velocity of the reduced particle, which conveniently describes the
binary, be $v=v_{\rm orb}$, and $\sigma$ be the velocity dispersion in
the cluster. An encounter of a soft binary ($v_{\rm orb} \ll \sigma$)
with another cluster system (in this gedanken experiment with the same
mass as the reduced particle) leads to the reduced particle gaining
kinetic energy, at the expense of the field star.  This increases $v$
and thus reduces $\left|e_{\rm b}\right|$, further increasing the
cross section for additional encounters.  Since the binding energy of
soft binaries is usually feeble, the extra energy input through the
encounter typically leaves the soft binary unbound, and the
corresponding cooling of the field is negligible. Similarly, for a
hard binary, $v_{\rm orb} \gg \sigma$, and an encounter reduces the
kinetic energy of the reduced particle, thus increasing $\left|e_{\rm
b}\right|$ of the binary. The field star leaves with an increased
kinetic energy.  Heggie (1975) and Hills (1975) provide, respectively,
detailed mathematical and numerical analysis of the cross sections for
these processes and arrive at the {\it Heggie--Hills law for
stimulated evolution}, namely that in a cluster, soft binaries get
softer, whereas hard binaries harden (see also Hut 1983).

Thus, disruption of soft binaries can cool the cluster, and the
hardening of hard binaries heats the cluster. The details depend
sensitively on the number ratio of hard to soft binaries, and,
assuming universal initial dynamical properties, ultimately on the
binding energy of the cluster. Cooling of a cluster was first noted in
the Trapezium-Cluster computations of KPM, and it turns out that the
most active cooling sources are those with $e_{\rm k} \simless
\left|e_{\rm b}\right| \simless 100\,e_{\rm k}$ (KII).

\noindent
{\bf Early quasi-equilibrium:} During the earliest phase (typically
shorter than a few initial crossing times, $t_{\rm cr}$), the cluster
expands as a result of mass segregation aided by the immediate onset
of some binary hardening (for the relative importance of both
processes see KII), which causes a decay of the velocity dispersion in
the expanding cluster. At the same time, the softest binaries are
disrupted until the cutoff, $e_{\rm b,cut}$, in the binary-star
binding-energy distribution satisfies $\left|e_{\rm b,cut}\right| >
e_{\rm k}$. At this critical ('thermal') time $t_t$, the cluster's
evolution changes by a reduced expansion rate.  More details are
available in KI.

The cutoff orbital period in a cluster thus indicates the maximum
concentration the cluster has ever had.

\noindent
{\bf Binary-disruption time-scale:} The time-scale for the disruption
of binaries in a young cluster is $t_{\rm cr}$, because it takes a few
global crossing times for most systems to have crossed at least once
through the central, dense region. This is demonstrated in fig.~2 of
KI, where $f_{\rm tot}(t)$ is plotted for clusters with $N = 800,
3000, 10000$ stars but the same $t_{\rm cr}$. At the same time,
$f_{\rm P}$ and $f_{\rm q}$ evolve through disruption of the systems
with smallest binding energy (i.e. large $P$ and small $q$). An
important empirical example is the ONC which has an $f_{\rm P}$ that
is depleted at $P\simgreat10^7$~days (Scally, Clarke \& McCaughrean
1999), which may be a result of binary disruption. This was predicted
to be the case by Kroupa (1995c, K3), who also suggests that $f_{\rm
q}$ should be significantly depleted at small mass ratios in the
Trapezium Cluster. BD--BD and BD--star systems are disrupted
efficiently because of their small $\left|e_{\rm b}\right|$, so that
typically $f_{\rm BD}\approx0.2$ after a few $t_{\rm cr}$ (fig.~6 in
Kroupa 2000c).

Should an observer catch a cluster before it is many $t_{\rm cr}$ old,
a {\it radially increasing binary proportion} would be evident, with a
decrease in the outer cluster regions as a result of the transient
halo of predominantly single stars forming as a result of the
encounters near the restless core. Such a radial signature of youth
persists until the cluster is well mixed, but the expanding
binary-depleted halo remains (KPM). It would be a dramatic discovery
if such a radial dependency of $f_{\rm tot}(r)$ would be found in the
ONC.

\noindent
{\bf Ejection of stars:} A cluster looses stars as a result of many
weak two-body encounters that elevate them above the escape energy
(e.g. Lee \& Goodman 1995). Rare, very close fly-bys of single stars
in a cluster potential can produce relatively large kinetic energy
changes and thus ejection events, leaving the other star more bound to
the cluster. Analytical estimates for idealised clusters (single,
equal-mass stars) show that the ejection events are about~4 times less
likely than losses through evaporation (Binney \& Tremaine 1987).

However, in realistic young clusters that have many primordial
binaries, three-body encounters (binary--'single', or
binary--very-hard binary) are not rare, because the cross section is
comparable to the semi-major axes of the binaries, which span a large
range, typically up to the 'thermal cutoff' given by the velocity
dispersion in the relaxed cluster.  Such events can lead to the binary
hardening and the 'single'-star receding with the kinetic energy
surplus gained from the hardened binary. The ejected star may be the
initial companion of the binary, if the companion is less massive than
the incoming system (e.g. fig.~6 in Hills 1975; Hills 1977).
Conservation of linear momentum implies that the binary recoils with a
corresponding velocity in the other direction to that of the 'single'
star, and if the gain in binding energy of the binary was sufficient,
this binary can also leave the cluster. An observer finds two systems
on opposite sides of a young cluster. Examples of such an event are
probably the two early~B stars lying on opposite sides of the
Monoceros~R2 cluster (Carpenter et al. 1997), as well as the two
early~B stars straddling the embedded cluster S~255-IR (Zinnecker,
McCaughrean \& Wilking 1993). In both cases, at least one of the
B~stars must be a binary.

Binary--binary interactions are also not rare in realistic clusters,
and, if the encountering binaries have similar $e_{\rm b}$, can lead
to complex behaviour, with temporarily bound but unstable four- or
three-body systems usually decaying into one binary and two single
stars.  After a time-lag, which is typically a few times the dynamical
time of the chaotic small-$N$ subsystem, the result can be ejections
with velocities up to a few~10~km/s (Sterzik \& Durisen 1998).  The
four-body systems usually decay in two steps, with one single star
being ejected first, followed later by the other single star. Since
the first ejection event is likely to be less energetic than the
second, which includes an already hardened system, and is not likely
to be in the same direction as the first ejection event because the
system is chaotic, the observer may find three systems receding away
from the cluster with different velocities, with at least one being a
binary.  Since the second decay is likely to release more energy, the
alignment would be such that the two more massive (i.e. brighter)
'stars' (one being the hardened binary) lie {\it nearly} co-linearly
on opposite sides of the cluster, and a third, less massive system
(from the first ejection event) typically lying closer to the cluster
but at some angle relative to the other two. An example of this may be
the possible multiple ejection event that lead to the run-away stars
AE~Aurigae, $\mu$~Columbae and the eccentric binary $\iota$~Orionis
(Hoogerwerf, de Bruijne \& de Zeeuw 2000). An example of an extreme
runaway $5\,M_\odot$ star is HIP~60350, which can also be traced back
to a cluster (Tenjes et al. 2000).

Run-aways can, however, also be produced as sling-shot events in
supernova explosions (Portegies Zwart 2000a), so that it is important
to build-up statistics from $N$-body calculations to seek possible
specific characteristics of run-aways produced by both mechanisms (see
also Leonard 1995).

The kinematical and dynamical properties of run-aways can be used to
constrain the binary populations and morphological properties of the
parent clusters (Clarke \& Pringle 1992).  Thus, binary-rich clusters
produce significantly more high kinetic-energy stars than single-star
clusters (fig.~5 in K3). Initially more concentrated clusters also
eject more high-energy stars. The highest velocities achieved in these
clusters ($N_{\rm bin}=200$ or $N_{\rm sing}=400$) are 40~km/s, but
rare, star-grazing encounters can expel stars at a few~100~km/s
(Leonard 1991). There are well-defined correlations between binary
proportion, $P$ and $m_{\rm sys}$ with ejection velocity. This
demonstrates that the dynamical properties retain a memory of the
dynamical events, which will be useful when interpreting the
distribution of young stars around sf regions on a statistical basis
(Leonard \& Duncan 1990; Sterzik \& Durisen 1998; Kroupa 1998).  Such
data will become available with the upcoming astrometry satellites
DIVA (R\"oser 1999) and GAIA (Gilmore et al. 1998).

\begin{figure}
\plotfiddle{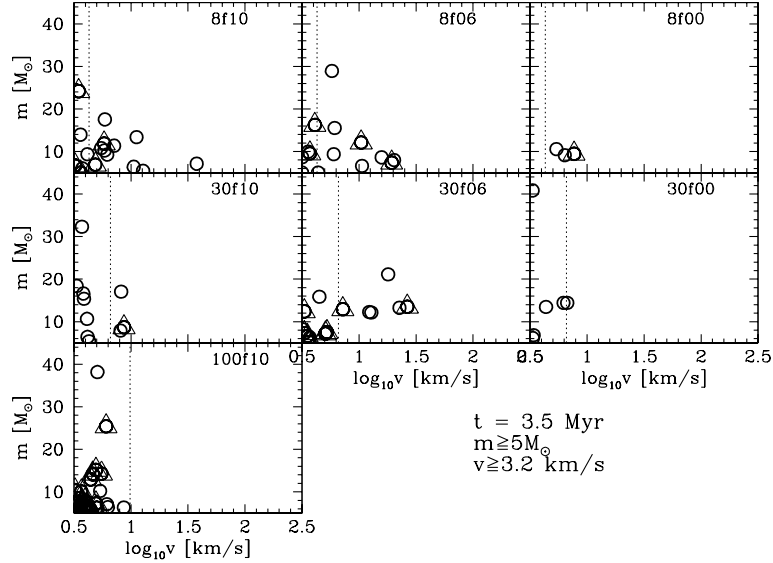}{7cm}{-90}{40}{40}{-150}{230}
\caption{\small Velocities of ejected massive stars at $t=3.5$~Myr for
the clusters discussed in KI ($N=800, 3000$ and~10000, from top to
bottom, with $f_{\rm tot}=1, 0.6$ and~0, from left to right; note that
all clusters here have the same IMF).  Triangles indicate
binaries. The top three panels contain data from $N_{\rm run}=10$
$N$-body calculations, for the middle three panels $N_{\rm run}=5$,
and for the bottom panel $N_{\rm run}=2$. Thus, the number of points
in the bottom panel should be divided by two to mentally normalise
the~7 data sets.  The vertical dotted lines indicate the escape
velocity, $v_{\rm esc}$, at $t=0$ from the cluster centre. The first
supernova explodes at $t\approx 4.5$~Myr.}
\label{fig:velOB}
\end{figure}

\begin{figure}
\plotfiddle{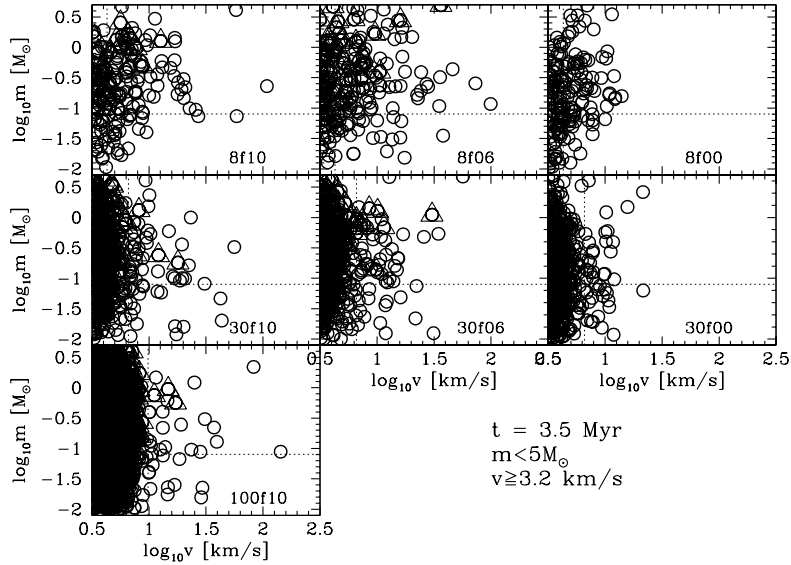}{7cm}{-90}{40}{40}{-150}{230}
\caption{\small As Fig.~\ref{fig:velOB}, but for low-mass stars.
The horizontal dotted lines indicate the brown dwarfs.}
\label{fig:vellm}
\end{figure}

Massive stars are also ejected, mostly through violent interactions in
the cluster core.  The velocities of pre-supernova dynamically ejected
stars from the cluster models of KI (central density as in the ONC, no
initial mass segregation and random pairing from the IMF) are
displayed in Figs.~\ref{fig:velOB} and~\ref{fig:vellm}.  These figures
demonstrate that the binary proportion among the ejected stars is very
low, as expected, and that stars with $m\simgreat5\,M_\odot$ do not
achieve $v\simgreat 40$~km/s in {\it any} of the models
by~3.5~Myr. Also evident is that the $N=10000$ cluster ejects no
massive stars within~3.5~Myr, and a small number of low-mass
stars. This is a result of the larger number of massive stars in this
model, the massive-star sub-population forming the core thus having a
longer relaxation time (i.e. being 'less collisional').  That the
primordial binary population has a significant effect on the ejection
rate is also apparent, the $f_{\rm tot}=1$ clusters having a
K2-period-distribution, whereas the $f_{\rm tot}=0.6$ models have a
log-normal period distribution, and thus a larger binding energy per
binary. Barely any ejections occur if $f_{\rm tot}=0$.
%-------------------------------------
% Figs: NB_EVAL/Strasb/vel_OBst.sm
%-------------------------------------
%-------------------------------------
% Figs: NB_EVAL/Strasb/vel_lmst.sm
%-------------------------------------

One immediate but still preliminary interpretation of these results is
that OB stars probably {\it must} form in energetic binaries (small
$P$ and large $q\le1$), and probably already near the centre of their
parent cluster, to explain the observed high-velocity OB run-aways,
thus tentatively supporting the formation scenario of Bonnell, Bate \&
Zinnecker (1998).

At later times (5--50~Myr), fig.~6 in KI shows that between
10~and~50~per cent of all surviving OB stars lie at distances larger
than $2\,R_{\rm tid}$ from their parent cluster, which has a tidal
radius $R_{\rm tid}$ (this includes companions flung out when their
primary explodes).

%======================================================================
\section{Forbidden Orbits -- A Window to Stellar Evolution}
\label{sec:e_p}
A binary involved in an energetic interaction, in which a companion
may be exchanged, usually ends up hardened and having a large $e$, and
possibly a fast space motion.  If $e$ is so large that the peri-astron
separation, $R_{\rm peri}$, is smaller than the sum of the stellar
radii, $R$, then the stars merge to form a rejuvenated, and if
massive, possibly an exotic star that may have been ejected from its
cluster (Leonard 1995; Portegies Zwart 2000b).  If the eccentricity is
not large enough for immediate physical collision, but $R_{\rm peri}$
is a few times $R$ but smaller than some critical value $R_{\rm
coll}$, then the binary orbit will tidally circularise rapidly, the
stars merging by the time the system is observed. If $R>R_{\rm coll}$
then the binary will circularise without merging within the age of the
system. If the system hardens or softens while circularising depends
on the angular momentum transfer between the stellar spins and orbit.

\begin{figure}
\plotfiddle{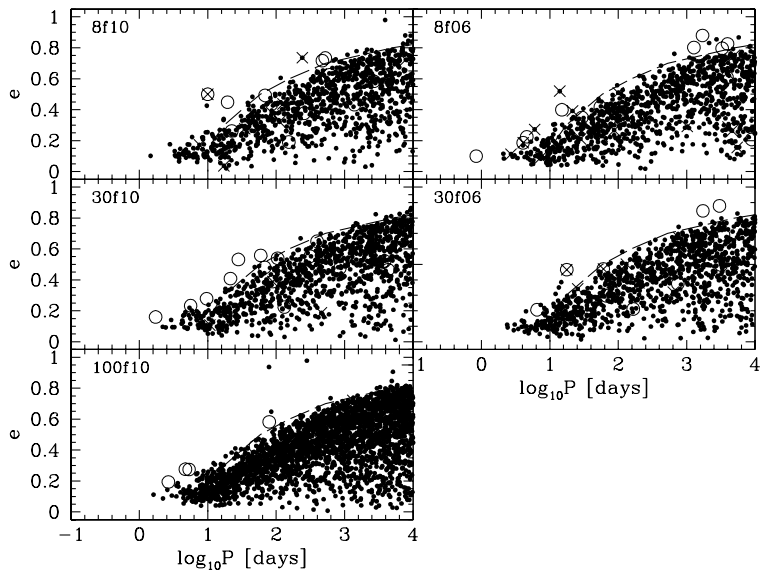}{6.6cm}{-90}{40}{40}{-150}{222}
\caption{\small Eccentricity--period diagram at $t=3.5$~Myr for the
clusters with $f_{\rm tot}=1.0$ and~0.6 shown in Figs.~\ref{fig:velOB}
and~\ref{fig:vellm} (no binaries form in the displayed region in the
clusters with $f_{\rm tot}=0$). Open circles have a system mass
$m\ge8\,M_\odot$, filled circles have $m<8\,M_\odot$; crosses show
systems with velocities $v>v_{\rm esc}=4.3, 6.6, 9.8$~km/sec (top to
bottom panel).  The long-dashed line delineates {\it forbidden orbits}
(on its left) from 'normal' binaries (K2), and is taken from Duquennoy
\& Mayor (1991).  Forbidden orbits are not eigenevolved once they are
produced (cf. fig.~5 in K2).}
\label{fig:e_p}
\end{figure}

Such eccentric systems can appear in the $e-{\rm log}_{10}P$ diagram
in a region usually avoided by binaries as a result of eigenevolution
(short $P$ and large $e$), and are thus referred to as {\it forbidden
orbits} (K2, fig.~5). A few candidate systems may exist (see K2), and
one particularly interesting example is Gl~586A, which has
$P=890$~days and $e=0.975$. It is discussed at length by Goldman \&
Mazeh (1994) as a prime example of how to test various tidal
circularisation theories. The rate of production of such orbits
depends on the number of stars in the cluster and on its
concentration, and is therefore greatest during the first few crossing
times. Fig.~3 in KI shows the average number of forbidden orbits per
cluster as a function of time, and Fig.~\ref{fig:e_p} plots the
corresponding $e-{\rm log}_{10}P$ diagrams. Note that not all
forbidden orbits have large centre-of-mass velocities. 

%-------------------------------------
% Figs: NB_EVAL/Strasb/ecc_per.sm
%-------------------------------------

Tidal circularisation depends on the coupling between the velocity
gradient (shear) in the stellar envelope due to the tide, and the
turbulent viscosity, and thus on the internal constitution of the
star. It is very efficient for fully convective stars, because 
the depth of the convection zone is essentially the handle that
connects the outer regions of the star with its interior, but the
theories for tidal circularisation are very uncertain. 

Zahn \& Bouchet (1989) show that most of the circularisation of
bloated fully convective pre-main sequence binaries occurs within the
Hayashi phase ($10^5-10^6$~yr). A longer evolution times-scale obtains
for older systems, and the longest period which is circular may be
useful as a clock to infer the age (or inversely the circularisation
time-scale if the age is known) of the system (see various
contributions in Duquennoy \& Mayor 1992).  Goldman \& Mazeh (1994)
study the circularisation of binaries with initially $e\approx1$, and
find that the time-scales for changes in semi-major axis, $a$, and
$e$, are $\tau_{\rm a}\equiv a/\dot{a} \propto
P^{16/3}\left(1-e\right)^{15/2}$ and $\tau_{\rm e}\equiv e/\dot{e}
\propto P^{16/3}\left(1-e\right)^{13/2}$. For $e=0.5,0.8$,
$\left(1-e\right)^{15/2}=0.0055, 5.7\times10^{-6}$ and
$\left(1-e\right)^{13/2}=0.011, 2.9\times10^{-5}$, respectively,
demonstrating that the time-scales are likely to be very short for
'forbidden systems' with large $e$.

Since forbidden systems have a short period ($P\simless 10^3$~days),
changes in $e$ and $P$ may be directly observable over a few to many
periods, in which case tidal circularisation theory may be testable
{\it if} the odd forbidden binary can be found that experienced an
encounter recently.  It may be worth re-checking the orbital
parameters of, for example, EK~Cep and P2486 (table~A2 in Mathieu
1994), and of the high-proper-motion system G253-44, which has
$P=19.38$~days, $e=0.52$ (Mazeh, Mayor \& Latham 1997).  Concerning
the ages of forbidden systems such as above mentioned Gl~586A, it is
worth keeping in mind that the time the system had for circularisation
is not the nuclear age of the stars, since the encounter that produced
the large $e$ may have happened long after the birth of the stars or
system.

%==============================================================
\vskip 3mm 
\noindent
%{\small This work was supported by DFG grant KR1635 at the Institute
%for Theoretical Astrophysics, University of Heidelberg, and made use
%of Aarseth's (1999) {\sc Nbody5} and {\sc Nbody6}.
{\small This work was supported by DFG grant KR1635, and made use
of Aarseth's (1999) {\sc Nbody5} and {\sc Nbody6}.

%==============================================================

%==============================================================

\end{document}